\documentclass[multphys,vecphys]{svmult}

\usepackage{makeidx}     
\usepackage{graphicx}    
\usepackage{multicol}    

\makeindex             

\begin{document}

\def\etal{et al. }
\def\apj{ApJ}
\def\apjl{ApJ\ Lett.}
\def\apjs{ApJ\ Suppl.}
\def\aas{A\&A Supp.}
\def\aa{A\&A}
\def\aal{A\&A Lett.}
\def\mnras{MNRAS}
\def\mnrasl{MNRAS Lett.}
\def\inpress{in press}
\def\inprep{in prep.}
\def\submit{submitted}
\def\mjysr{MJy/sr }
\def\inu{{I_{\nu}}}
\def\fnu{{F_{\nu}}}
\def\bnu{{B_{\nu}}}
\def\mic{$\mu$m}

\title*{New insights on the thermal dust from the far-infrared to the centimeter}
\titlerunning{New insights on the thermal dust}

\author{X. Dupac\inst{1} \and J.-P. Bernard\inst{2} \and N. Boudet\inst{2}
\and M. Giard\inst{2} \and J.-M. Lamarre\inst{3} \and C. M\'eny\inst{2} \and F. Pajot\inst{4} \and I. Ristorcelli\inst{2}}
\authorrunning{Dupac et al.}

\institute{European Space Agency - ESTEC, Astrophysics Division, Keplerlaan 1, 2201 AZ Noordwijk, The Netherlands,
\texttt{xdupac@rssd.esa.int}
\and CESR, 9 av. du Colonel Roche, BP 4346, 31028 Toulouse cedex 4, France
\and LERMA, Obs. de Paris, 61 av. de l'Observatoire, 75014 Paris, France
\and IAS, Campus d'Orsay, b\^at. 121, 15 rue Clemenceau, 91405 Orsay cedex, France
}
%
%
\maketitle

\begin{abstract}
We present a compilation of PRONAOS-based results concerning the temperature dependence of the dust submillimeter spectral index, including data from Galactic cirrus, star-forming regions, dust associated to a young stellar object, and a spiral galaxy.
We observe large variations of the spectral index (from 0.8 to 2.4) in a wide range of temperatures (11 to 80 K).
These spectral index variations follow a hyperbolic-shaped function of the temperature, high spectral indices (1.6-2.4) being observed in cold regions (11-20 K) while low indices (0.8-1.6) are observed in warm regions (35-80 K).
Three distinct effects may play a role in this temperature dependence: one is that the grain sizes change in dense environments, another is that the chemical composition of the grains is not the same in different environments, a third one is that there is an intrinsic dependence of the dust spectral index on the temperature due to quantum processes.
This last effect is backed up by laboratory measurements and could be the dominant one.

We also briefly present a joint analysis of WMAP dust data together with COBE/DIRBE and COBE/FIRAS data.
\end{abstract}

\section{Introduction}
To accurately characterize dust emissivity properties represents a major challenge of nowadays astronomy.
In the submillimeter domain, large grains at thermal equilibrium (e.g. \cite{desert90}) dominate the dust emission.
This thermal dust is characterized by a temperature and a spectral dependence of the emissivity which is usually simply modelled by a spectral index.
The temperature, density and opacity of a molecular cloud are key parameters which control the structure and evolution of the clumps, and therefore, star formation.
The spectral index ($\beta$) of a given dust grain population is directly linked to the internal physical mechanisms and the chemical nature of the grains.

It is generally admitted from Kramers-K\"onig relations that 1 is a lower limit for the spectral index.
$\beta$ = 2 is particularly invoked for isotropic crystalline grains, amorphous silicates or graphitic grains.
However, it is not the case for amorphous carbon, which is thought to have a spectral index equal to 1.
Spectral indices above 2 may exist, according to several laboratory measurements on grain analogs.
Observations of the diffuse interstellar medium at large scales favour $\beta$ around 2 (e.g. \cite{boulanger96}, \cite{dunne01}).
In the case of molecular clouds, spectral indices are usually found to be between 1.5 and 2.
However, low values (0.2-1.4) of the spectral index have been observed in circumstellar environments, as well as in molecular cloud cores.

\section{The dust spectral index as seen by PRONAOS\label{pro}}

PRONAOS (PROgramme NAtional d'Observations Submillim\'etriques) is a French
balloon-borne submillimeter experiment \cite{ristorcelli98}.
Its effective wavelengths are 200, 260, 360 and 580 \mic, and the angular resolutions are 2$'$ in bands 1 and 2, 2.5$'$ in band 3 and 3.5$'$ in band 4.
The data analyzed here were obtained during the second flight of PRONAOS in September 1996, at Fort Sumner, New Mexico.
This experiment has observed various phases of the interstellar medium, from diffuse clouds in Polaris \cite{bernard99} and Taurus \cite{stepnik03} to massive star-forming regions in Orion (\cite{ristorcelli98}, \cite{dupac01}), Messier 17 \cite{dupac02}, Cygnus B, and the dusty envelope surrounding the young massive star GH2O 092.67+03.07 in NCS.
The $\rho$ Ophiuchi low-mass star-forming region has also been observed, as well as the edge-on spiral galaxy NGC 891 \cite{dupac03b}.

We fit a modified black body law to the spectra: $\inu = \epsilon_0 \; \bnu(\lambda,T) \; (\lambda/\lambda_0)^{-\beta}$, where $\inu$ is the spectral intensity (MJy/sr), $\epsilon_0$ is the emissivity at $\lambda_0$ of the observed dust column density, $\bnu$ is the Planck function, $T$ is the temperature and $\beta$ is the spectral index.
In most of the areas, we use either only PRONAOS data or PRONAOS + IRAS 100 \mic~data.
We restrain the analysis to all fully independent (3.5$'$ side) pixels for which both relative errors on the temperature and the spectral index are less than 20\%.
This procedure allows to reduce the degeneracy effect between the temperature and the spectral index.
Dupac et al. (2001, 2002) have shown by fitting simulated data that this artificial anticorrelation effect was small compared to the effect observed in the data.

\begin{figure}[!ht]
\includegraphics[scale=.6]{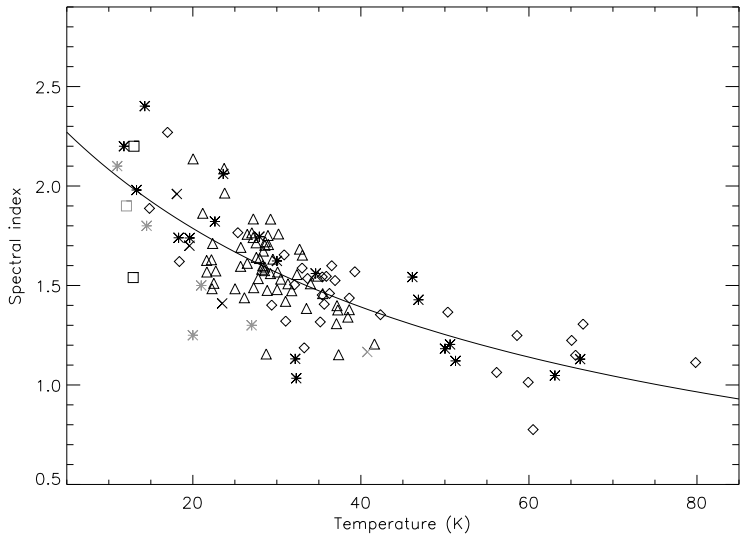}
\caption[]{Spectral index versus temperature, for fully independent pixels in Orion (black asterisks), M17 (diamonds), Cygnus (triangles), $\rho$ Ophiuchi (grey asterisks), Polaris (black squares), Taurus (grey square), NCS (grey cross) and NGC 891 (black crosses).
The full line is the result of the best hyperbolic fit: $\beta = {1 \over 0.4 + 0.008 T}$
}
\label{compil}
\end{figure}

We present in Fig. \ref{compil} the spectral index - temperature relation observed.
The temperature in this data set ranges from 11 to 80 K, and
the spectral index also exhibits large variations from 0.8 to 2.4.
One can observe an anticorrelation on these plots between the temperature and the spectral index, in the sense that the cold regions have high spectral indices around 2, and warmer regions have spectral indices below 1.5.
In particular, no data points with $T >$ 35 K and $\beta >$ 1.6 can be found, nor points with $T <$ 20 K and  $\beta <$ 1.5.
This anticorrelation effect is present for all objects in which we observe a large range of temperatures, namely Orion, M17, Cygnus and $\rho$ Ophiuchi.
It is also remarkable that the few points from other regions are well compatible with this general anticorrelation trend.
The temperature dependence of the emissivity spectral index is well fitted by a hyperbolic approximating function.

Several interpretations are possible for this effect: one is that the grain sizes change in dense environments, another is that the chemical composition of the grains is not the same in different environments and that this correlates to the temperature, a third one is that there is an intrinsic dependence of the spectral index on the temperature, due to quantum processes such as two-level tunneling effects.
Additional modeling, as well as additional laboratory measurements and astrophysical observations, are required in order to discriminate between these different interpretations.
More details about this analysis and the possible interpretations can be found in \cite{dupac03a}.

\section{The Galactic dust emission as seen by COBE and WMAP}

\begin{figure}[!ht]
\includegraphics[scale=.6]{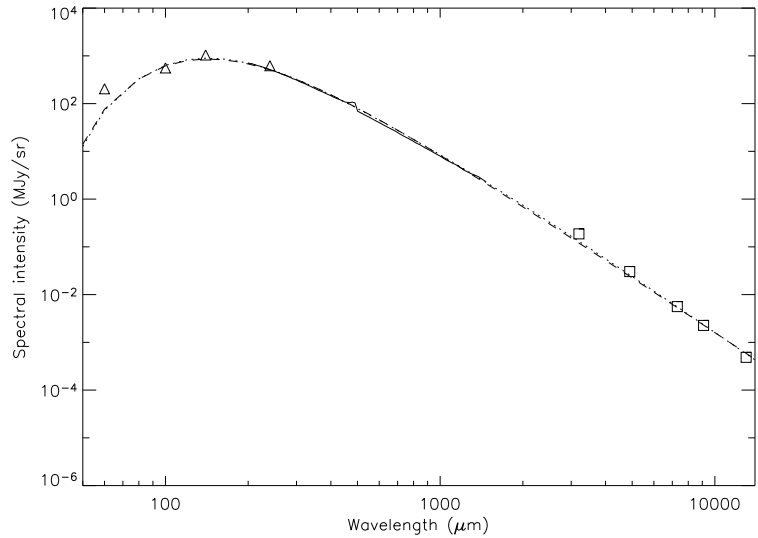}
\caption[]{Galactic dust spectrum in the Galactic center region.
DIRBE data points are presented as triangles, the FIRAS spectrum is presented as a full line, and the squares are the WMAP data points.
The dashed and dotted lines are the results of, respectively, the one-component and two-component fits, without the 60 \mic~data points.
}
\label{sp}
\end{figure}

We combine data sets from COBE (DIRBE data points and FIRAS spectrum) and WMAP in order to investigate the Galactic dust spectral energy distribution on the whole sky.
We convolve all data to the FIRAS 7$^o$ angular resolution, in order to obtain a consistent data set.
This is done by co-adding higher resolution data from DIRBE and WMAP into 7 degree-sided pixels.
The resulting spectrum is shown in Fig. \ref{sp} for the Galactic center area.
It is well fitted by a 21 K, $\beta$=1.86 one-dust component model.
These values are well compatible with the PRONAOS temperature - spectral index relation (see Section \ref{pro}).
More results and conclusions from this COBE - WMAP joint analysis will soon be available.

\printindex

\begin{thebibliography}{99.}

\bibitem{bernard99} Bernard, J.-P., Abergel, A., Ristorcelli, I., et al.: 1999, A\&A, 347, 640

\bibitem{boulanger96} Boulanger, F., Abergel, A., Bernard, J.-P., et al.: 1996, A\&A, 312, 256

\bibitem{desert90} D\'esert, F.-X., Boulanger, F., Puget, J.-L.: 1990, A\&A, 237, 215

\bibitem{dunne01} Dunne, L., Eales, S.A.: 2001, MNRAS, 327, 697

\bibitem{dupac01} Dupac, X., Giard, M., Bernard, J.-P., et al.: 2001, \apj, 553, 604

\bibitem{dupac02} Dupac, X., Giard, M., Bernard, J.-P., et al.: 2002, \aa, 392, 691

\bibitem{dupac03a} Dupac, X., Bernard, J.-P., Boudet, N., et al.: 2003, \aal, 404, L11

\bibitem{dupac03b} Dupac, X., del Burgo, C., Bernard, J.-P., et al.: 2003, MNRAS, 344, 105

\bibitem{ristorcelli98} {Ristorcelli}, I., {Serra},
  G., {Lamarre}, J.-M., et al.: 1998, ApJ, 496, 267

\bibitem{stepnik03} Stepnik, B., Abergel, A., Bernard, J.-P., et al.: 2003, A\&A, 398, 551

\end{thebibliography}
\end{document}